# Advanced Optics Experiments Using Nonuniform Aperture Functions

Lowell T. Wood

Department of Physics

University of Houston

Houston, TX 77204-5005

**ABSTRACT**

A method to create instructive, nonuniform aperture functions using spatial frequency filtering is described. The diffraction from a single slit in the Fresnel limit and the interference from a double slit in the Fraunhofer limit are spatially filtered to create electric field distributions across an aperture to produce apodization, inverse apodization or super-resolution, and apertures with phase shifts across their widths. The diffraction effects from these aperture functions are measured and calculated. The excellent agreement between the experimental results and the calculated results makes the experiment ideal for use in an advanced undergraduate or graduate optics laboratory to illustrate experimentally several effects in Fourier optics.

## I. INTRODUCTION

Most intermediate and advanced optics books discuss apodization[1-12], the reduction in amplitude of the secondary maxima in diffracted light. This reduction in amplitude is accomplished by creating an aperture function whose transmission is smaller at the edges than in the center of the aperture. Aperture functions are generally first encountered when the study of Fourier optics is begun. Calculations using aperture functions such as the cosine function and the linear taper are common in optics books. Diffraction from such aperture functions are characterized by an increased width of the central diffraction peak and a reduction in the amplitude of the secondary maxima relative to that of the central diffraction peak. Inverse apodization or super-resolution is discussed by Goodman[10] and by Wilson[11]. Wilson[12] also provides some excellent figures showing the results of apodization, super-resolution, and complex aperture functions with $\pi$ and $\pi/2$ phase shifts across the apertures, all in two dimensions. Jacquinot and Roizen-Dossier[13] give a very detailed overview of apodization.

Although such calculations are numerous, few experiments have been described that make it possible for students to study these effects experimentally. In fact, only Palmer[7] and Ramsay et al.[14], describe experiments suitable for advanced optics laboratories. In this paper, a method for performing such experiments is described, and results are compared to calculations.

## II. THEORY

Consider the one-dimensional Fraunhofer diffraction integral given by

$$F(x) = \int_{-\infty}^{+\infty} A(x') e^{\frac{ikxx'}{R}} dx', \qquad (1)$$

where $x$ is the position in the observation plane, $x'$ is the position in the aperture plane, $k$ is the propagation number of the incident plane wave, and $R$ is the distance from the aperture plane to the observation plane. The integral is recognized as the Fourier transform $F(x)$ of the aperture function $A(x')$ if $\omega$, the spatial frequency, is defined to be $kx/R$. The limits of integration are contained in $A(x')$ and extend from $-b/2$ to $+b/2$, where $b$ is the width of the aperture. It is straightforward to carry out the integration for many aperture functions that satisfy the Fraunhofer diffraction condition. What is not discussed in optics books is an experimental technique to create such aperture functions and to study their diffraction patterns. In the next section, a method to create nontrivial and interesting aperture functions is described, and results from the experiments and calculations are provided.

### III. EXPERIMENTAL METHODS

The experimental arrangement for creating the aperture functions and measuring their diffraction patterns is shown in Figure 1. Light from a spatially filtered, 0.9-mW helium-neon laser transmits through a double slit having slit widths of 0.04 mm and a center-to-center slit separation of 0.50 mm and produces a well-known diffraction pattern. A detailed view of the central portion of the irradiance is shown in Figure 2. The 0.25-mm wide single slit acts as a filtering slit that is designed to transmit only that portion of the diffraction pattern that produces the desired aperture function. It is important to remember that the pattern measured by a detector is the irradiance, whereas the electric field is present at the plane of the filtering slit. A 500-mm focal length, anti-reflection coated cylindrical lens is placed after the filtering slit, and the line camera is located on the back focal plane of the lens. (The line camera was a Thorlabs model LC-1 with a pixel spacing of 7 micrometers and a 12-bit A/D converter to measure irradiances.) Observing the irradiance on the back focal plane of the lens ensures that the Fraunhofer diffraction limit is maintained. (Because of the small Fresnel number, 0.2, there was, however, no measureable difference in the patterns with and without the lens in place.)

The filtering slit remains a constant width of 0.25 mm, and its distance from the double slit is adjusted so that only the desired portion of the pattern transmits and creates the aperture function within the boundaries of the filtering slit. From the double slit, three aperture functions are created, named as follows: the full-sine aperture function, the cosine aperture function, and the half-sine aperture function. From Figure 2, region *abcde* creates the full-sine aperture function, region *cde* gives the cosine aperture function, and region *bcd* provides the half-sine aperture function. The inverted cosine aperture function is created by a single-slit diffraction pattern in the Fresnel limit, as shown in Figure 3. Only the central minimum is transmitted, and the aperture function is approximated by $A(x') = 1 - \varepsilon \cos(\pi x'/b)$ for $-b/2 < x' < b/2$ and $A(x') = 0$ elsewhere. Here, $\varepsilon$ is 0.38 since the irradiance is 0.38 at the central maximum, and its square root, 0.62, is the amplitude of the electric field.

It should be noted that the regions selected using the double slit are not exactly sine and cosine functions because of the modulation by the single-slit diffraction. However, by choosing widely-separated, narrow slits, the approximation is quite sufficient for this experiment and has the advantage of making the diffraction integrals of the aperture functions relatively straightforward to do. The fit for the inverted cosine aperture function is not nearly as good as the other fits, but the agreement between experimental results and calculated results seems to indicate that the exact dependence of the aperture function on position is not nearly as important as ensuring that the transmission in the center is smaller than that at the edges.

## IV. RESULTS AND DISCUSSION

Figure 4 shows both the calculated and the measured normalized irradiances for each aperture function considered here: the uniform aperture function, cosine aperture function, inverted cosine aperture function, full-sine aperture function, and half-sine aperture function. The dashed lines are the calculated curves, and the solid curves are the measured curves. The two vertical dotted lines represent the first zeroes of the uniform aperture function irradiance for ease of comparison. Clearly, there is excellent agreement between the measured and calculated results. The bottom curve shows the graph of the experimental and calculated irradiances for the uniform aperture function, *i.e.*, the usual single slit, for comparison. The cosine aperture function irradiance clearly shows the effects of apodization, *i.e.,* broadening of the central diffraction peak and relative reduction in strength of the secondary maxima. The inverted cosine aperture function irradiance, on the other hand, has its central diffraction peak reduced in width with the attendant increase in the relative strength of the secondary maxima; *i.e.*, it shows inverse apodization or super-resolution.

The irradiance from the full-sine aperture function and that of the half-sine aperture function illustrate similar effects but with a $\pi$ phase shift across the aperture. The full-sine aperture function has zero amplitude at the boundaries with larger amplitudes within the aperture, although not at the center. The zero amplitude at the boundaries causes the secondary maxima to be reduced in strength, whereas the width of the central peak is broadened compared to that of the half-sine aperture function. The $\pi$ phase shift causes the aperture to take on the characteristics of two apertures, each with half the width of the physical aperture, thereby producing a double slit with each slit having an electric field distribution $\pi$ phase shifted with respect to the other. The two dashed lines near the top of the graph represent the width of the irradiance pattern from a uniform aperture 0.125 mm wide. Therefore, two interference fringes are visible within the central diffraction peak, causing a minimum to be formed at the center. The half-sine aperture behaves similarly, but the absolute value of the electric field is maximum at the edges of the aperture and zero at the center. As expected, the half-sine aperture has relatively larger secondary maxima and a narrower central diffraction peak. The Table gives a summary of the most important features for the irradiance of the aperture functions, expressed in terms of the spatial frequency $\omega$. The irradiance functions are analytical expressions, as are the zero-to-zero central irradiance widths, $\Delta\omega$, with the exception of the inverted cosine function that was determined numerically from the graph. The values $\Delta x$ given below the $\Delta\omega$'s are the zero-to-zero central irradiance widths for the parameters used in this experiment. Each irradiance ratio is the ratio of the first secondary irradiance maximum to the maximum irradiance observed, and all are determined numerically from the graphs.

## V. CONCLUSIONS

A method for creating instructive, nonuniform aperture functions has been demonstrated, and diffraction effects from each aperture function have been measured and calculated. Both the calculated results and the measured results confirm some of the basic results of Fourier optics, *i.e.*, apodization, inverse apodization or super-resolution, and diffraction from apertures with phase shifts across them. The ease of producing such aperture functions and the excellent agreement between the measured and calculated values make this experiment ideal for the advanced undergraduate or graduate optics laboratory. The author has used parts of this experiment in an advanced undergraduate optics laboratory for several years with good results. As an additional note, such techniques may have a place in the general problem of optical beam shaping, as the number of aperture functions is limited only by one's imagination.

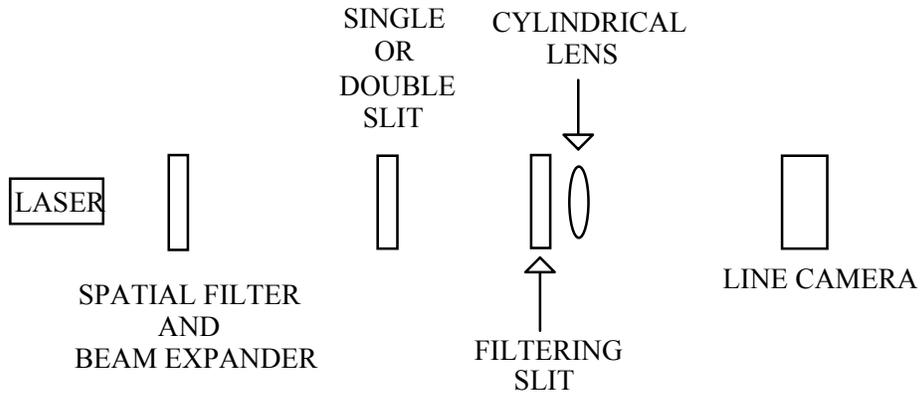

Figure 1. Experimental arrangement showing the spatial filter and beam expander for the laser, the single or double slit to create the aperture function, the filtering slit to transmit the appropriate portion of the diffraction/interference pattern, and the line camera for measuring the diffraction caused by the aperture function.

FIGURE 1 – WOOD

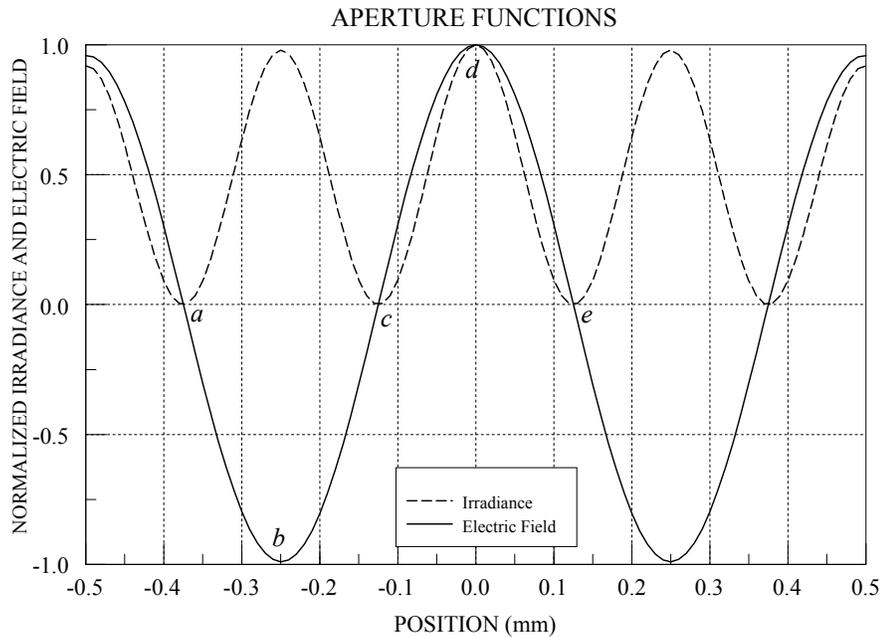

Figure 2. Creating the aperture functions. The distance from the double slit to the filtering slit is 197±1 mm. The half-sine aperture function, *b* to *c* to *d*, and the cosine aperture function, *c* to *d* to *e*, both span the 0.25-mm width of the filtering slit. Creating the full-sine aperture function, *a* to *b* to *c* to *d* to *e*, requires moving the filtering slit to one-half the distance given above.

FIGURE 2 – WOOD

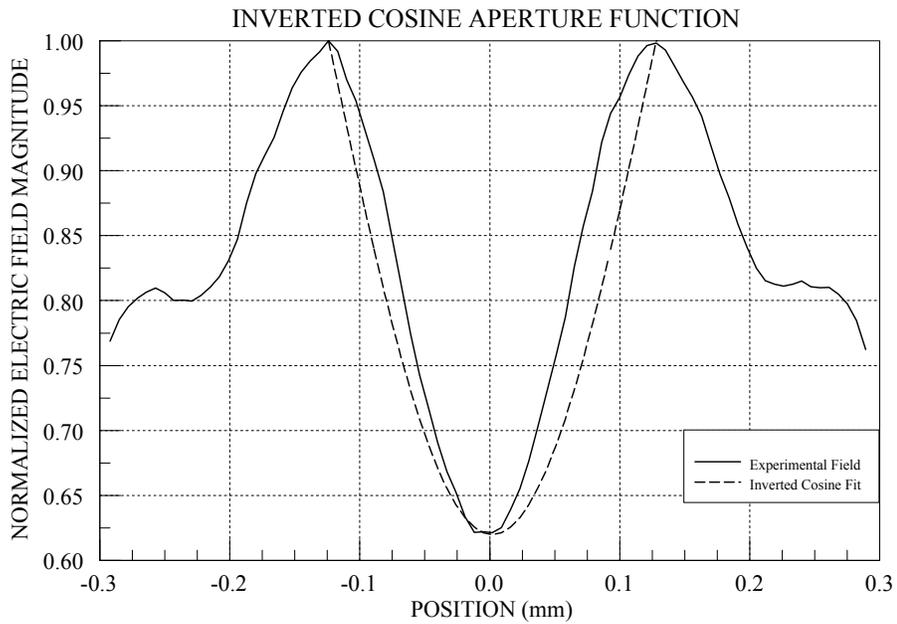

Figure 3. The inverted cosine aperture function was created using a single slit 0.80 mm wide with the 0.25-mm wide filtering slit located 156±1 mm away from it. The parameters were selected to produce a minimum field that is 62% of the maximum field. The fitted curve is given by $A(x) = 1 - \varepsilon \cos(\pi x/b)$ for $-b/2 < x < b/2$ and $A(x) = 0$ elsewhere, with $b = 0.25$ mm and $\varepsilon = 0.38$.

FIGURE 3 – WOOD

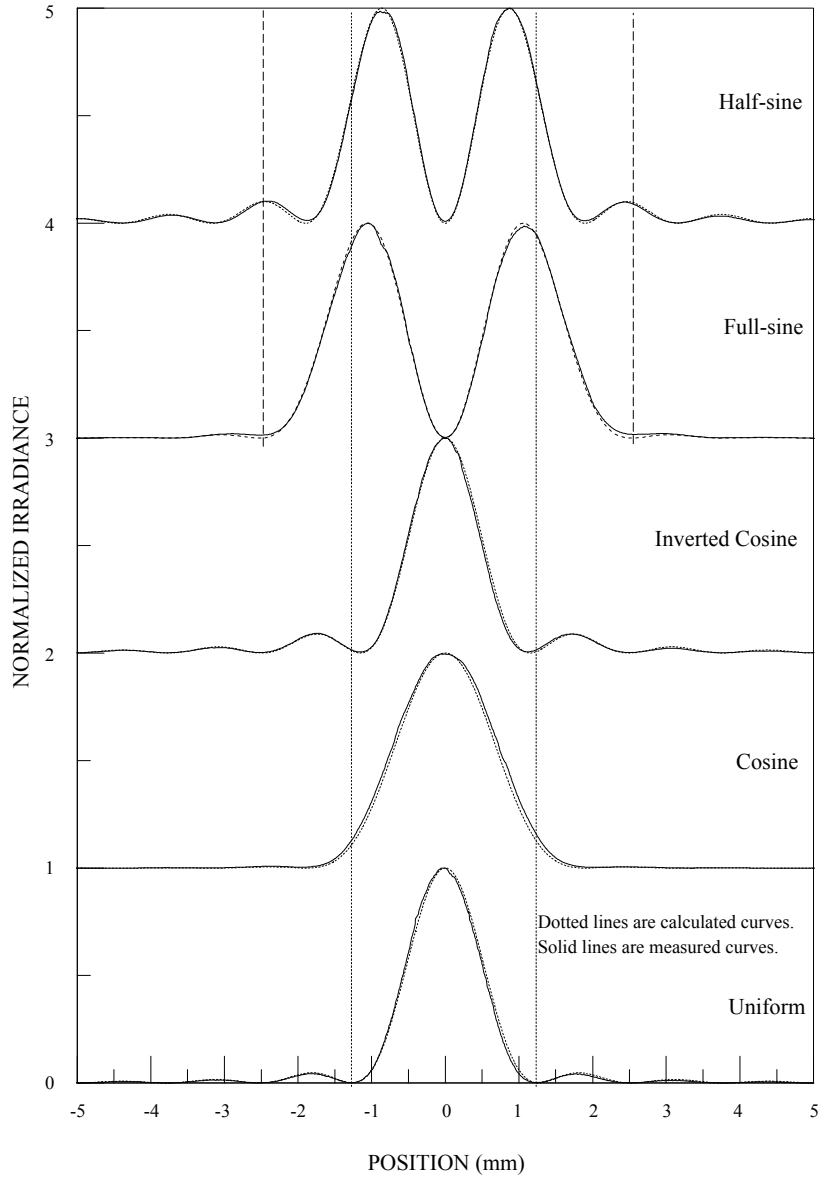

Figure 4. Normalized irradiance for each of the aperture functions measured. Each curve is normalized to one and displaced by one vertical unit so that the irradiance from each aperture function is shown on the same graph for comparison. The dotted vertical lines are the first zeroes of the uniform aperture function irradiance. The dashed vertical lines at the top of the figure are the first irradiance zeroes for a uniform aperture having a width of 0.125 mm. (See discussion in text.)

FIGURE 4 – WOOD

| Aperture Function | Irradiance on Diffraction Plane | Zero-to-zero Central Irradiance Width | Irradiance Ratio (numerical) | Aperture Function Name |
|---|---|---|---|---|
| $A(x) = 1$ for $-b/2 < x < b/2$ $A(x) = 0$ elsewhere | $I(\omega) = \left[ \dfrac{b \sin\left(\dfrac{\omega b}{2}\right)}{\left(\dfrac{\omega b}{2}\right)} \right]^2$ | $\Delta\omega = \left(\dfrac{4\pi}{b}\right)$ $\Delta x = 2.53$ mm | 0.047 | Uniform |
| $A(x) = \sin(\pi x/b)$ for $-b/2 < x < b/2$ $A(x) = 0$ elsewhere | $I(\omega) = \left[ \dfrac{2b^2 \omega \cos\left(\dfrac{\omega b}{2}\right)}{\left(\pi^2 - \omega^2 b^2\right)} \right]^2$ | $\Delta\omega = \left(\dfrac{6\pi}{b}\right)$ $\Delta x = 3.80$ mm | 0.100 | Half-sine |
| $A(x) = \sin(2\pi x/b)$ for $-b/2 < x < b/2$ $A(x) = 0$ elsewhere | $I(\omega) = \left[ \dfrac{4\pi b \sin\left(\dfrac{\omega b}{2}\right)}{\left(4\pi^2 - \omega^2 b^2\right)} \right]^2$ | $\Delta\omega = \left(\dfrac{8\pi}{b}\right)$ $\Delta x = 5.06$ mm | 0.015 | Full-sine |
| $A(x) = \cos(\pi x/b)$ for $-b/2 < x < b/2$ $A(x) = 0$ elsewhere | $I(\omega) = \left[ \dfrac{2\pi b \cos\left(\dfrac{\omega b}{2}\right)}{\left(\pi^2 - \omega^2 b^2\right)} \right]^2$ | $\Delta\omega = \left(\dfrac{6\pi}{b}\right)$ $\Delta x = 3.80$ mm | 0.0050 | Cosine |
| $A(x) = 1 - \varepsilon \cos(\pi x/b)$ for $-b/2 < x < b/2$ $A(x) = 0$ elsewhere | $I(\omega) = \left[ \dfrac{2\pi b \varepsilon \cos\left(\dfrac{\omega b}{2}\right)}{\left(\omega^2 b^2 - \pi^2\right)} + \dfrac{b \sin\left(\dfrac{\omega b}{2}\right)}{\dfrac{\omega b}{2}} \right]^2$ | $\Delta\omega = \left(\dfrac{3.64\pi}{b}\right)$ $\Delta x = 2.30$ mm (numerical) | 0.088 | Inverted Cosine |

Table. Summary of several features of the irradiances for each aperture function. The central irradiance widths for the parameters used in this experiment are calculated by using the result $k\Delta x/R = \Delta\omega$. $\Delta x$ values have uncertainties of ±0.03 mm.

TABLE - WOOD